\documentclass[aps,superscriptaddress,double-spaced,floatfix,showkeys,showpacs]{revtex4}
\usepackage{graphicx,amsmath,bbm,bm,array,subfigure}
\usepackage{appendix}
\usepackage{lipsum}
\usepackage{dsfont}
\usepackage{calrsfs}
\usepackage[linktocpage=true,colorlinks=true,linkcolor=blue,citecolor=blue,urlcolor=cyan]{hyperref}
\usepackage{MnSymbol}
\def\nn{\nonumber\\}

\newcommand\Snn{\mathcal{S}_{nn}(\vec{k},\omega)}
\newcommand\K{\mathcal{K}}
\begin{document}
\title {Dynamical spectral structure of density fluctuation near QCD critical point}
\author{Md Hasanujjaman}
\email{jaman.mdh@gmail.com}
\affiliation{Department of Physics, Darjeeling Government College, Darjeeling- 734101, India}
\author{Golam Sarwar}
\email{golamsarwar1990@gmail.com}
\affiliation{Discipline of Physics, School of Basic Sciences,
Indian Institute of Technology, Indore, India} 
\author{Mahfuzur Rahaman}
\email{mahfuzurrahaman01@gmail.com }
\affiliation{Variable Energy Cyclotron Centre, 1/AF Bidhan Nagar, Kolkata- 700064, India}
\affiliation{Homi Bhabha National Institute, Training School Complex, Mumbai - 400085, India}
\author{Abhijit Bhattacharyya}
\email{abhattacharyyacu@gmail.com}
\affiliation{Department of Physics,University of Calcutta, 92, A.P.C. Road, Kolkata-700009, India}
\author{Jan-e Alam}
\email{jane@vecc.gov.in}
\affiliation{Variable Energy Cyclotron Centre, 1/AF Bidhan Nagar, Kolkata- 700064, India}
\affiliation{Homi Bhabha National Institute, Training School Complex, Mumbai - 400085, India}
\def\zbf#1{{\bf {#1}}}
\def\bfm#1{\mbox{\boldmath $#1$}}
\def\hf{\frac{1}{2}}
\def\sl{\hspace{-0.15cm}/}
\def\omit#1{_{\!\rlap{$\scriptscriptstyle \backslash$}
{\scriptscriptstyle #1}}}
\def\vec#1{\mathchoice
        {\mbox{\boldmath $#1$}}
        {\mbox{\boldmath $#1$}}
        {\mbox{\boldmath $\scriptstyle #1$}}
        {\mbox{\boldmath $\scriptscriptstyle #1$}}
}
\def \beq{\begin{equation}}
\def \eeq{\end{equation}}
\def \beqa{\begin{eqnarray}}
\def \eeqa{\end{eqnarray}}
\def \pd{\partial}
\def \nn{\nonumber}
\begin{abstract}
The expression for the dynamical spectral structure of the density fluctuation near 
the QCD critical point has been derived using linear response theory within the purview 
of Israel-Stewart relativistic viscous hydrodynamics. The change in spectral structure of 
the  system as it moves toward critical end point has been studied. The effects of the 
critical point have been introduced in the system  through a realistic equation of state 
and the scaling behaviour of various transport coefficients and thermodynamic response 
functions.  We have found that the  Brillouin and the  Rayleigh peaks are distinctly 
visible when the system is away from critical point but the peaks tend to merge near 
the critical point. The sensitivity of structure of the spectral function on wave vector ($k$)  
of the sound wave has been demonstrated. It has been shown that the Brillouin peaks get merged
with the Rayleigh peak because of the absorption of sound waves in the vicinity of the critical 
point.
 \end{abstract}
\pacs{12.38.Mh, 12.39.-x, 11.30.Rd, 11.30.Er}
\maketitle
\section{ Introduction}
\label{intro}
Relativistic Heavy Ion Collisions experiments (RHIC-E) are 
aimed at exploring properties of high temperature~\cite{Cabibbo:1975ig} 
and high density~\cite{Collins:1974ky} state of strongly interacting matter. 
The degrees of freedom of hadronic matter under extreme conditions of temperature and
density are deconfined quarks and gluons and their interaction is governed by Quantum Chromodynamics (QCD).  
The study of the deconfined state of strongly interacting matter is relevant for
understanding the evolution of micro-second old universe, composition of the core of the
compact astrophysical objects ({\it e.g.} neutron star) 
and properties of non-abelian gauge theory in medium~\cite{qgp1,qgp2}. 
The RHIC-E  provide opportunities to verify different theoretical predictions 
on the thermal nature of QCD matter~\cite{ Busza:2018rrf}. 
Results of these experiments indicate that after collision 
of two heavy ions at relativistic energies, a strongly interacting perfect 
fluid medium consisting of quarks and gluons -
called quark gluon plasma (QGP)~\cite{ShuryakPhysRep} - is formed~\cite{npa2005}. 
The fluid with high internal pressure expands hydrodynamically,  cools consequently and revert to hadronic phase. 
The thermal properties of the QGP
can be described by two independent variables, the temperature ($T$) and baryonic chemical potential ($\mu$) associated with the
conservation of net baryon number in the system. 
The values of $T$ and $\mu$ of the system depends on the energy of collisions of the nuclei.
Numerical simulation of QCD in the high $T$ and low $\mu$ $(\rightarrow 0)$ 
region predict that the quark-hadron transition is a cross-over 
~\cite{Fodor:2001pe,Asakawa:1989bq, Halasz:1998qr, deForcrand:2002hgr, Aoki:2006we}. 
However, calculations based on several QCD inspired models indicate that at high $\mu$ region 
the transition is first order~\cite{deForcrand:2002hgr, Endrodi:2011gv}.  
Therefore, it is expected that between the cross over and first order transitions there exists a region 
or point in $T-\mu$ plane
where the first order transition ends and crossover begins~\cite{Fodor:2004nz}. 
This point is the QCD Critical End  Point(CEP).
At present, lattice QCD results for high  $\mu$ is not available   
due to the well known sign problem for spin 1/2 particles.  As a 
result the prediction on the precise position of  
the CEP is not possible from first principle calculations~\cite{Gavai:2014ela}.
Due to the lack of first principle calculations 
there have been a large numbers of effective field theory based studies on 
the QCD phase at non-zero baryon density 
~\cite{PNJL10,PNJL11,PNJL12,PQM1,PQM2,PQM3}. 
These model studies indicate the existence of CEP in the phase diagram. 
However, the position of the 
CEP is still ambiguous as its location depends on the parameters of the models used.
 
Different points of the QCD phase diagram in the $\mu-T$ plane can be reached experimentally by 
varying the energy of the colliding nuclei. The
deconfined, {\it i.e} the QGP systems with different $T$ and $\mu$ can be produced 
by colliding heavy ions with different energies. These
systems will follow different trajectories in $\mu-T$ plane while
making a transition from QGP to hadrons due to cooling caused by expansion. 
Parallel to the theoretical endeavor, experimental efforts are on to explore these
trajectories passing through/near the CEP by varying beam energy~\cite{Aggarwal:2010cw}. 

It is well-known that  the density-density correlation length diverges at the critical point. 
The effects of the divergence 
on the baryon number fluctuations, particle correlations and correlations of density  
fluctuations will have better chance to be detected provided the fluctuation survives the 
hadronic evolution. 
It is important to understand the correlations of these fluctuations theoretically  
for identifying signatures of CEP in data. The fluid dynamical descriptions of the system at the
CEP breaks down. However, it is possible to identify a region near the CEP where
the fluid dynamics remains valid and can be used to understand the properties of 
the system~\cite{Stanley}. The effects of CEP on the evolution of matter 
goes as input through the equation of state (EoS)  and via the critical behaviour of the 
transport coefficients and response functions of the fluid.

The correlation of density fluctuations can be investigated through
the spectral structure ($\Snn$) in Fourier space near the CEP.
Spectral function has been studied experimentally
in the condensed matter physics laboratories to estimate 
the speed of sound by  using scattering of light. 
The position of Brillouin peaks (B-Peaks) in the $\Snn$  is
connected to the speed of sound and various transport coefficients like
thermal conductivity, shear and bulk viscosities and response functions
like specific heats. Sadly, such external probes are not available to
examine the properties of QCD matter near CEP.

The spectral structure has been estimated in Ref.~\cite{kunihiro} 
but the effects of EoS containing the critical point have been ignored.  It has been shown
in that work that the EoS  plays a vital role in determining the behaviour of $\Snn$, 
especially its strength at the Rayleigh peak (R-peak)
changes by several orders of magnitude if EoS with effects of CEP is incorporated.
The EoS has strong effects on B-peak too. 
It is also very important to understand whether all the hydrodynamic modes travel with the 
same speed or not.  The R-peak and the B-peaks will be closer for slower modes  
even at points away from the CEP. Therefore, the structure of $\Snn$ will shed 
light on the speed of the perturbation propagating as sound wave.

The paper is organized as follows. The EoS containing the effects of CEP is discussed
in the next section. The expression for $\Snn$ has been
derived in section III within the scope of the Israel-Stewart (IS) relativistic hydrodynamics
in Eckart frame of reference. Section IV is devoted to discuss the critical behaviour
of some transport coefficients and response functions. 
Results are presented in section VI and section VII is dedicated to summary and discussions. 

\section{Equation of State}
\label{QCDEoS}
One of our main objectives is to investigate the role of the CEP on 
the density fluctuation. In order to observe it, we have constructed an 
EoS ~\cite{mitedu,nonaka,hasan} which contains the effects of CEP. 
To make the article self-contained, we here briefly discuss the construction of the EoS. 
The universality hypothesis suggests that the 3D Ising model and
the QCD belong to the same universality class. Therefore, we can map the calculation of 3D Ising 
model onto the QCD phase diagram. It is well-known that magnetization $(M)$ of Ising model 
is analogous to a critical entropy density ($s_c$) in QCD. 
We can express $M$ at any temperature $(T)$ 
as a function of reduced temperature, $r=(T-T_c)/T_c$ and applied magnetic field 
$(\mathcal{H})$, where $T_c$ is the critical temperature, with the location of CEP  
at $(r, \mathcal{H})=(0, 0)$. Thus, $r<0$ represents the first order phase transition  
and $r>0$ indicates the crossover transition. A linear mapping between CEP of Ising model and 
CEP of QCD phase diagram is performed by assuming a critical region by the following relation	 
	\begin{equation}
	r=\frac{T-T_c}{\Delta T_c}, \,\,\,\, \mathcal{H}=\frac{\mu-\mu_c}{\Delta \mu_c}
	\end{equation} 
	where $(\mu_c,T_c)$ represents the location of the CEP
	in the QCD phase diagram.  
	The extension of the critical region can be determined by the values of
	$\Delta T_c$ and $\Delta \mu_c$ where $\Delta T_c$ and $\Delta \mu_c$ are chosen as 
	extensions of the critical region along T and $\mu $ axis respectively.  
	The critical entropy density can be expressed as 
	\begin{eqnarray}
	s_c= \frac{M (r,\mathcal{H})}{\Delta T_c}= M\Big (\frac{T- T_c}{\Delta T_c}, \frac{\mu- \mu_{c}}{\Delta \mu_{c}} \Big ) \frac{1}{\Delta T_c}
	\label{eq2}
	\end{eqnarray}
	In order to construct the EoS, we first construct a  dimensionless entropy density  as
	\begin{eqnarray}
	S_c= A (\Delta T_c, \Delta \mu_{c})s_c (T, \mu)
	\label{eq3}
	\end{eqnarray}
when $A $ is defined as 
	\begin{equation}
	A(\Delta T_{c},\Delta \mu_{c}) = B \sqrt{\Delta T^2_{c}+\Delta \mu_{c}^2)}
	\label{eq4}
	\end{equation}
	Here $B$ is also a dimensionless quantity which 
represents the elongation of the critical region. 
	In this work, we use  
	$(T_c, \mu_c)= (154 \text{MeV}, 367 \text{MeV})$ 
	with $(\Delta T_{c}, \Delta \mu_{c},B)=(0.1\,\text{GeV}, 0.2\, \text{GeV}, 2)$.
	The construction of the entropy density is done by making  a correspondence between the 
entropy density of QGP ($s_Q$) and the hadronic ($s_H$) phases with use of $S_c$ as a switching 
function. The final result reads  
	\begin{eqnarray}
	S (T, \mu)=\frac{1}{2}[1-{\rm tanh} \ S_c (T,\mu)] s_{Q}(T,\mu) + \frac{1}{2}[1+{\rm tanh} \ S_c (T,\mu)] s_H(T,\mu)
	\label{eq5}
	\end{eqnarray}
	where we calculate $s_{Q}$  as\cite{satarov}
	\begin{eqnarray}
	s_{Q}(T,\mu)=\frac{32+21N_{f}}{45}\pi^{2}T^{3}+\frac{N_{f}}{9}\mu^{2}T 
	\label{eq6}
	\end{eqnarray}
	with $N_{f}$ being  the number of flavor of quarks.\\The hadronic entropy density
	($s_H$) can be estimated from the following expression ~\cite{pbraun,Sarwar}, 
\begin{eqnarray}
s_H(T,\mu)=\pm \sum_{i}\frac{g_{i}}{2\pi^{2}}
\int^{\infty}_{0}dp^{\prime}{p^\prime}^{2}\Big[ln\Big(1\pm \{exp(E_{i}-\mu_{i})/T\}\Big)\pm \frac{E_{i}-\mu_{i}}{T\{exp(E_{i}-\mu_{i})/T\pm 1\}}\Big]
\label{eq7}
\end{eqnarray}
where the sum extends over all hadrons with mass up to 2.5 GeV~\cite{Sarwar},
	$g_i$   represents the statistical degeneracy factor
	and $E_{i}=\sqrt{{p^\prime}^{2}_{i}+m^{2}_{i}}$ is the energy of the $i^{\text{th}}$ hadron.
	
	Once we know the entropy density, the thermodynamic quantities {\it e.g.} 
baryon number density, pressure 
	and energy density can be evaluated as follows. 
The net baryon number density ($n$) is given by 
	\begin{eqnarray}
	n(T, \mu)= \int_{0}^{T} \frac{\partial S (T^{'}, \mu)}{\partial \mu} dT^{'}
	\label{eq8}
	\end{eqnarray}
	The pressure can be estimated as 
	\begin{eqnarray}
	p(T, \mu)=  \int_{0}^{T} S (T^{'}, \mu) dT^{'} 
	\label{eq10}
	\end{eqnarray}
	and finally, the energy density is given by, 
	\begin{eqnarray}
	\epsilon(T, \mu)= Ts(T,\mu)-p(T,\mu)+\mu n
	\label{eq11}
	\end{eqnarray}
	To get the first order phase boundary,  the discontinuity 
	in the entropy density along the transition line also need to be considered. We add the following term 
	to the Eq.\ref{eq8} to  take into account  this possibility (for $T>T_{c}$) 
	\begin{eqnarray}
	\Big|\frac{\partial T_{c}(\mu)}{\partial \mu}\Big|\Big[S(T_c(\mu)+\delta,\mu)-S(T_c(\mu)-\delta,\mu)\Big]
	\label{Tslope}
	\end{eqnarray}
	where $\Big|\frac{\partial T_{c}}{\partial \mu}\Big|=tan\theta_{c}$, is the tangent 
	at the $T_c$ and $\delta$ is the small deviation in  temperature   from $T_{c}$.
	The value of $T_{c}$  for the first order transition depends on $\mu$ 
	as indicated in Eq.~\ref{Tslope}. 
\section{Brief review on the relativistic viscous fluid dynamics: 
Israel-Stewart (IS) theory in Eckart frame}
In this work we use  the IS relativistic hydrodynamic theory to describe the fluid \cite{Israel}
dynamics in the Eckart frame of reference. 
It was developed earlier by Muller in the context of non-relativistic fluids \cite{Muller}. 
Israel-Stewart method generalizes the standard model of the entropy current for out of equilibrium systems, and then enforces the second law of thermodynamics in the simplest possible way.
The baryonic chemical potential, $\mu$, is introduced for
the conservation of the net baryon number, $n$. 
In general there are two choice of frames of references, namely Landau-Lifshitz (LL)
~\cite{Landau} and Eckart~\cite{Eckart} frames. 
The LL frame represents a local rest frame where the energy dissipation is zero but the net number dissipation 
(diffusion) is non-zero. Whereas the Eckart frame represents a local rest frame where the net charge dissipation 
is vanishing but the energy dissipation is non-vanishing. We adopt the Eckart frame of reference 
where the particle current can be written as:
\beqa
N^{\mu}=nu^{\mu}
\label{eq9}
\eeqa
where, the metric convention is $g^{\mu \nu}=(-1,1,1,1)$. 
The conserved baryon number density is given by $ n=-u_{\mu}N^{\mu}$. 
The energy momentum tensor is given by
\beqa
T^{\mu \nu}= \epsilon u^{\mu}u^{\nu}+(p+\Pi)\Delta^{\mu \nu}+q^{\mu}u^{\nu}+q^{\nu}u^{\mu}+\pi^{\mu \nu}
\label{eq10}
\eeqa
where $\epsilon$ is energy density, $p$ is the pressure, $\Delta^{\mu \nu}=g^{\mu \nu}+u^{\mu}u^{\nu}$
is the projection operator. It has the properties $\Delta^{\mu\nu}u_{\mu}=\Delta^{\mu\nu}u_{\nu}=0$ and $\Delta^{\mu\nu}\Delta^{\alpha}_{\nu}=\Delta^{\mu\alpha}.$
In the above equation $\Pi$ and $\pi^{\mu \nu}$ represent the trace part (bulk pressure) and the symmetric transverse traceless part 
(shear pressure) of the symmetric viscosity tensor respectively and $q^{\mu}$ is the heat flow vector. They describe the out of equilibrium properties of the fluid and they satisfy the constraints
\beqa
u_{\mu}q^{\mu}=u_{\mu}\pi^{\mu\nu}=\pi^{\nu\mu}=\pi^{\mu\nu}=\pi^{\mu}_{\mu}=0
\eeqa
As the Eckart frame is characterized by no charge flow, the four velocity is expressed as 
$u^{\mu}=N^{\mu}/\sqrt{-N^{\alpha}N_{\alpha}}$. The Muller-Israel-Stewart theory 
takes into account of 
general relativity with the method of Grad's 14-moment approximation and successfully 
restores the causality condition for the relativistic viscous fluids 
by considering 
second order gradients of hydrodynamic variables. The general forms of 
$q^{\mu}, \Pi, \pi^{\mu \nu}$ which 
contain additional coefficients arising due to inclusion of second order gradients are \cite{Hiscock,Biro,Baier}:
\beqa
\Pi&=& -\zeta[\pd_{\mu}u^{\mu}+\beta_{0}D\Pi-\alpha_{0}\pd_{\mu}q^{\mu}] \nn\\
\pi^{\lambda \mu}&=&-2\eta \Delta^{\lambda\mu\alpha\beta}\Big[\partial_{\alpha}u_{\beta}+\beta_{2}D\pi_{\alpha\beta}-\alpha_{1}\partial_{\alpha}q_{\beta}\Big]\nn\\
q^{\lambda}&=&\kappa T\Delta^{\lambda\mu} [-\frac{1}{T}\partial_\mu T -Du_{\mu}-\beta_1 D{q_\mu}+\alpha_0\partial_\mu \Pi +\alpha_1\pd_{\nu}\pi ^{\nu}_{\mu} ] 
\label{eq11}
\eeqa
where, $D\equiv u^\mu\partial_\mu$, is
known as co-moving derivative and in the local rest frame (LRF) $D\Pi =\dot{\Pi }$ 
represents the time derivative. Here $\eta$, $\zeta $, $\kappa$ are the coefficient of shear viscosity, bulk viscosity and thermal conductivity 
respectively. The double symmetric traceless projection operator is defined by 
$\Delta^{\mu\nu\alpha\beta}=
\frac{1}{2}\big[\Delta^{\mu\alpha}\Delta^{\nu\beta}+\Delta^{\mu\beta}\Delta^{\nu\alpha}-
\frac{2}{3}\Delta^{\mu\nu}\Delta^{\alpha\beta}\big]$ and $\beta _0,\beta_1,\beta_2$ are relaxation coefficients, 
 $\alpha_0$ and $\alpha_1$ are coupling coefficients. The relaxation times for the 
bulk pressure ($\tau_{\Pi}$), the heat flux ($\tau_q$) and the shear tensor ($\tau_{\pi}$) 
are defined as~\cite{muronga} 
\begin{equation}
\tau_{\Pi}=\zeta \beta_0, \,\,\,\,\tau_q=k_BT\beta_1,\,\,\,\, \tau _{\pi}=2\eta \beta_2
\label{eq12}
\end{equation}
The relaxation lengths which  couple to   heat flux and bulk  pressure 
($l_{\Pi q}, l_{q\Pi} $), the heat flux and shear tensor $(l_{q\pi}, l_{\pi q})$ 
are defined as follows:
\begin{equation}
l_{\Pi q}=\zeta \alpha_0,\,\,\,\, l_{q\Pi}=k_B T \alpha _0, \,\,\,\,l_{q\pi}=k_BT\alpha_1,\,\,\,\, 
l_{\pi q}=2\eta \alpha_1 
\label{eq13} 
\end{equation}  
In the ultra-relativistic limit, $\beta(=m/T)\rightarrow 0$), where $m$ is the mass
of the particle. We also have~\cite{Israel},
\begin{eqnarray}
\alpha_0 \approx  6\beta^{-2}p^{-1},\,\,\,\, \alpha_1\approx -\frac{1}{4}p^{-1},\,\,\,\,
\beta_0 \approx 216 \beta^{-4} p^{-1},
\beta_1 \approx \frac{5}{4}p^{-1},\,\,\,\, \beta_2 \approx \frac{3}{4}p^{-1} 
\label{eq14}
\end{eqnarray}
The relativistic viscous fluid are described by the following two equations corresponding to the 
conservation of energy-momentum and the net charge (net baryon number here) density:
\beqa
\pd_{\mu}T^{\mu \nu}&=&0 \nn\\
\pd_{\mu}N^{\mu}&=&0 
\label{eq15}
\eeqa
together with second law of thermodynamics governed by,
\beqa
\pd_{\mu}S^{\mu}\ge 0
\label{eq16}
\eeqa
\subsection{Linearized Hydrodynamic equations}
The hydrodynamic Eqs.~\ref{eq15} are non-linear, partial differential equations
which are difficult (if not impossible) to solve analytically in general. Presently our
goal is to obtain the $\Snn$ of the dynamical density
fluctuations in wave vector-frequency space.  The hydrodynamical equations  mentioned
above can be linearized to describe  small perturbations in
thermodynamical variables (small deviations from the equilibrium values of
the variable). These linearized
equations can be solved to obtain the density fluctuations.
Let $Q_0$ ($Q$) denote a thermodynamic quantity in (away from) equilibrium. 
For small perturbation $\delta Q$, $Q$ can be written as: $Q=Q_{0}+\delta Q$ where
$Q$ can be any of the quantities among $n, \epsilon, u^{\alpha}, q^{\alpha}, s, 
\Pi, \pi^{\alpha \beta}$, etc ($v_0$ has been taken as zero here). 
The linearized hydrodynamic equations around the 
equilibrium~\cite{kunihiro,Romatchske,hasan,Sayantani,Grozdanov} become:
\beqa
0&=&\frac{\pd \delta n}{\pd t}+n_{0} \vec{\nabla}.\delta \vec{v}  \nn\\
0&=&h_{0}\frac{\pd \delta v}{\pd t}+\nabla(\delta p+ \delta \Pi)+\frac{\pd \delta q}{\pd t}+\vec{\nabla}.\delta\vec{\pi} \nn\\
0&=& \delta \Pi +\zeta[\vec{\nabla}.\delta \vec{v}+\beta_{0}\frac{\pd \delta \Pi}{\pd t}-\alpha_{0}\vec{\nabla}.\delta \vec{q}] \nn\\
0&=& \delta \pi^{ij}+\eta[\pd^{i}\delta v^{j}+\pd^{j}\delta v^{i}-\frac{2}{3}g^{ij}\vec{\nabla}.\delta \vec{v}+2\beta_{2}\frac{\pd \delta \pi^{ij}}{\pd t} -\alpha_{1}(\pd^{i}\delta q^{j}+\pd^{j}\delta q^{i}-\frac{2}{3}g^{ij}\vec{\nabla}.\delta \vec{q})\nn\\
0&=& \delta q-\kappa T_{0}[-\frac{\nabla \delta T}{T_{0}}- \frac{\pd \delta v}{\pd t}-\beta_{1}\frac{\pd \delta q}{\pd t}+\alpha_{0}\nabla \delta \Pi+\alpha_{1}\vec{\nabla}.\delta\vec{\pi}] \nn\\
0&=& n_{0}\frac{\pd \delta s}{\pd t}+\frac{1}{T_{0}}\vec{\nabla}.\delta \vec{q}
\label{eq17}
\eeqa
We decompose the fluid four velocity along the directions parallel and perpendicular to the 
direction of wave vector, $\vec{k}$ and call them $\delta \vec{v_{||}}$ and $\delta\vec{v_{\perp}}$ 
respectively, {\it i.e.} $\vec{k}.\delta \vec{v_{\perp}}=0$. The hydrodynamic equations 
can be solved for a given set of initial condition, $n(0), v_{||}(0), T(0), q(0), \Pi(0)$
and  $\pi(0)$, by using the Fourier-Laplace transformation as:

\beqa
\delta Q(\vec{k}, \omega)= \int^{\infty}_{-\infty} dr \int^{\infty}_{0}dt e^{-i(\vec{k}.\vec{r}-\omega t)} \delta Q(\vec{r}, t)
\label{eq18}
\eeqa
The $\delta p$ and $\delta s$ can be written in terms of the independent variables $n$ and $T$ as 
follows by using the thermodynamic relations:
\beqa
\delta p&=&\big(\frac{\pd p}{\pd n}\big)_{T}\delta n+ \big(\frac{\pd p}{\pd T}\big)_{n}\delta T\nn\\
\delta s&=&\big(\frac{\pd s}{\pd n}\big)_{T}\delta n+ \big(\frac{\pd s}{\pd T}\big)_{n}\delta T
\label{eq19}
\eeqa
We use Eqs. \ref{eq18} and \ref{eq19}
to write down the  longitudinal linearized hydrodynamic equation as: 
\beqa
\delta Q(\vec{k}, \omega)=\mathds{M} \delta Q(k,0)
\label{eq20}
\eeqa
where, 
\begin{widetext}
\beqa
\mathds{M} =
\begin{bmatrix}
    i\omega & ikn_{0} & 0 & 0 & 0  & 0   \\
    \frac{ik}{h_{0}} \big(\frac{\pd p}{\pd n}\big)_{T}& i\omega & \frac{ik}{h_{0}} \big(\frac{\pd p}{\pd T}\big)_{n}& \frac{i\omega}{h_{0}}  & \frac{ik}{h_{0}} & \frac{ik}{h_{0}}  \\
    0 & ik\zeta & 0 & -ik\alpha_{0}\zeta & 1+i\omega \beta_{0} \zeta & 0 \\
  
    0 & -i\frac{4}{3}k\eta & 0 & i\frac{4}{3}\alpha_{1}k \eta & 0 & 1+2i\omega \beta_{2}\eta\\
    0 & i\omega \kappa T_{0} & ik\kappa & 1+i\omega \beta_{1} \kappa T_{0} & ik\alpha_{0} \kappa T_{0} & ik \alpha_{1} \kappa T_{0} \\
      -i\omega n_{0}\big(\frac{\pd s}{\pd n}\big)_{T} & 0 &   i\omega n_{0}\big(\frac{\pd p}{\pd T}\big)_{n}&\frac{ik}{T_{0}}  & 0  & 0  \\
\end{bmatrix}
\label{eq21}
\eeqa
\beqa
\delta Q(\vec{k},\omega)=
\begin{bmatrix}
\delta n(\vec{k},\omega) \\
\delta v_{||}(\vec{k},\omega)\\
\delta \Pi(\vec{k},\omega)\\
\delta \pi_{||}(\vec{k},\omega)\\
\delta q_{||}(\vec{k},\omega)\\
\delta T(\vec{k},\omega)
\end{bmatrix}
;\hspace{0.2cm}
\delta Q(\vec{k},0)=
\begin{bmatrix}
\delta n(\vec{k},0) \\
\delta v_{||}(\vec{k},0)+\frac{1}{h_{0}}\delta q_{||}(\vec{k},0)\\
i\omega \beta_{0}\zeta \delta \Pi(\vec{k},0)\\
-2\beta \eta \delta \pi_{||}(\vec{k},0)\\
-\kappa T_{0}\delta v_{||}(\vec{k},0)+\kappa T_{0}\beta_{1}\delta q_{||}(\vec{k},0)\\

-n_{0}\big(\frac{\pd s}{\pd n}\big)_{T}\delta n(\vec{k},0)+n_{0}\big(\frac{\pd s}{\pd T}\big)_{n}\delta T(\vec{k},0)\\
\end{bmatrix}
\label{eq22}
\eeqa
We are concerned about the density fluctuation which is given by, 
\beqa
\delta n(\vec{k},\omega)&=&\Big[ \mathds{M}^{-1}_{11}-n_{0}\big(\frac{\pd s}{\pd n}\big)_{T}\mathds{M}^{-1}_{16}\Big ] \delta n(\vec{k},0)
+\Big[\mathds{M}^{-1}_{12} -\kappa T_{0}\mathds{M}^{-1}_{15}\Big]\vec{\delta v_{||}}(\vec{k},0) \nn\\
&&+\mathds{M}^{-1}_{13}\Big[ i\omega \beta_{0} \zeta\Big] \delta \Pi(\vec{k},0) 
 - \mathds{M}^{-1}_{14}\Big[ 2\beta_{2}\eta\Big]\delta\pi_{||}(\vec{k},0) \nn\\
&&\Big[\frac{1}{h_{0}}\mathds{M}^{-1}_{12}+\kappa T_{0}\beta_{1}\mathds{M}^{-1}_{15}\Big] \delta q_{||} (\vec{k},0)\Big]
+  \mathds{M}^{-1}_{16} \Big[ n_{0} \big(\frac{\pd s}{\pd T}\big)_{n}\Big]\delta T(\vec{k},0)
\label{eq23}
\eeqa
\end{widetext}
The $\mathcal{S^\prime}_{nn}(\vec{k},\omega)$ is calculated by using the following correlator:
\beqa
\mathcal{S^\prime}_{nn}(\vec{k},\omega)=\big< \delta n(\vec{k},\omega)\delta n(\vec{k},0)\big>
\label{eq24}
\eeqa
The correlation between two independent thermodynamic variables, say, $Q_i$ and $Q_j$  vanishes i.e
\beqa
\big< \delta Q_{i}(\vec{k},\omega)\delta Q_{j}(\vec{k},0)\big>=0
\label{eq25}
\eeqa
The required correlator, $\mathcal{S^\prime}_{nn}(\vec{k},\omega)$ is obtained as:
\beqa
\mathcal{S^\prime}_{nn}(\vec{k},\omega)&=&\Big[ \mathds{M}^{-1}_{11}-n_{0}\big(\frac{\pd s}{\pd n}\big)_{T}\mathds{M}^{-1}_{16}\Big ] \big< \delta n(\vec{k},0)\delta n(\vec{k},0)\big>
\eeqa
We define $\Snn$ as 
\begin{equation}
\mathcal{S}_{nn}(\vec{k},\omega)=\frac{\mathcal{S^\prime}_{nn}(\vec{k},\omega)}
   {\big< \delta n(\vec{k},0)\delta n(\vec{k},0)\big>}
\label{eq26}
\end{equation}
The variation of $\Snn$ with $k$ and $\omega$ for given values of transport coefficients
and other relevant thermodynamic variables are studied below. 
The full expression for $\Snn$ has been given in the appendix A.
In the small $k$ limit with 
$\alpha_0\rightarrow 0,\alpha_1\rightarrow 0,\beta_0\rightarrow 0,\beta_1\rightarrow 0, \beta_2\rightarrow 0$
the results for Navier-Stokes hydrodynamics can be recovered from the full expression given in the appendix A.
We will see below that the variation of $\Snn$ with $\omega$ 
admits three peaks positioned at $\omega =0$ and  $\omega=\pm\omega_B$.
The $\omega_B$ is a function of  $k$, velocity of sound and other thermodynamic 
variables. 
The peak at $\omega=0$ is called the Rayleigh(R)-peak and the doublet symmetrically situated at 
$\pm\omega_{B}$ are called Brillouin (B)-peaks. 
The quantities $\big(\frac{\pd p}{\pd n}\big)_{T} ,\big(\frac{\pd p}{\pd T}\big)_{n},
\big(\frac{\pd s}{\pd n}\big)_{T}, \big(\frac{\pd s}{\pd T}\big)_{n}$ 
appearing in $\Snn$ can be evaluated in terms of relevant thermodynamic variables 
(see Appendix). 
In  condensed matter physics the  $\Snn$ is measured by utilizing the relation
between intensity of light scattering with density-density correlation~\cite{Stanley}. However,
any such direct measurement of the corresponding critical opalescence in QCD is not possible.
The possibility of the measurement of QCD opalescence by measuring jet quenching has
been proposed~\cite{csorgo}.
 
\section{Behaviour of $\Snn$  near CEP}
The  dynamical structure function, $\Snn$ defined in Eq.~\ref{eq26}
depends on the transport coefficients,
$\eta, \zeta$ and  $\kappa$ as well as on the
four partial derivatives, $\big(\frac{\pd p}{\pd n}\big)_{T} ,\big(\frac{\pd p}{\pd T}\big)_{n}$, 
$\big(\frac{\pd s}{\pd n}\big)_{T}$ and $\big(\frac{\pd s}{\pd T}\big)_{n}$. 
It also depends on the coupling coefficients $(\alpha_{0}, \alpha_{1})$ and the relaxation 
coefficients $(\beta_{0}, \beta_{1}, \beta_{2})$ which infiltrate through  the 
IS hydrodynamics. The partial derivatives appearing in the 
$\Snn$ contain the effects of CEP through EoS. 
The behaviour of the transport coefficients and the response functions 
near the CEP is characterized by the critical exponents.
As the CEP is approached some of these quantities start to diverge.
A thermodynamic variable, $f(r)$ near the CEP can be written 
as~\cite{Stanley}:
\beqa
f(r)=Ae^{\lambda}(1+Be^{y}+...)
\label{eq31}
\eeqa
where, $y> 0$ and the critical exponent $\lambda$ can be defined as:
\beqa
\lambda=\lim_{r \to 0} \frac{ln f(r)}{ln (r)}
\label{eq32}
\eeqa
$\lambda$ can be either  +ve or -ve correspondingly
$f(r)$ will vanish or diverge at the CEP. 
Now the partial derivatives appearing in the expression for $\Snn$ can be expressed 
as (see Appendix): 
\beqa
\big(\frac{\pd p}{\pd n}\big)_{T}&=&\frac{1}{n_{0}\kappa_{T}}, \big(\frac{\pd p}{\pd T}\big)_{n}= \mu nc^{2}_{s}\alpha_{p} \frac{C_{v}}{C_{p}} \nn\\
 \big(\frac{\pd s}{\pd n}\big)_{T}&=& \frac{h_{0}c^{2}_{s}\alpha_{p}}{n_{0}\gamma} ,\big(\frac{\pd s}{\pd T}\big)_{n}=\frac{C_{v}}{T_{0}}
\label{eq35}
\eeqa
where, $\kappa_{T}=\frac{1}{n}\big(\frac{\pd n}{\pd p}\big)_{T}$, is the 
isothermal compressibility, 
$\alpha_{p}=-\frac{1}{n}\big(\frac{\pd n}{\pd T}\big)_{p}$ is volume 
expansivity coefficient, $C_p, C_v$ are specific heats at constant 
pressure and volume respectively and $\gamma=C_p/C_v$. 
The thermodynamic variables can be  expressed in 
terms of reduced temperature $r$ near the CEP as ~\cite{Kapusta,Guida}:
\beqa
&&\kappa_{T}=\kappa_{0}|r|^{-\gamma^\prime}, C_{v}=C_{0}|r|^{-\alpha}, C_{p}=\frac{\kappa_{0}T_{0}}{n_{0}}\big(\frac{\pd p}{\pd T}\big)^{2}_{n}|r|^{-\gamma^\prime},\nn\\
&& c^{2}_{s}=\frac{T_{0}}{n_{0}h_{0}C_{0}}\big(\frac{\pd p}{\pd T}\big)^{2}_{n}|r|^{\alpha}, \alpha_{p}=\kappa_{0}\big(\frac{\pd p}{\pd T}\big)_{n}|r|^{-\gamma^\prime} \nn\\
&& \eta=\eta_{0}|r|^{1+a_{k}/2-\gamma^\prime}, \zeta= \zeta_{0}|r|^{-\alpha}, 
\kappa=\kappa_{0}|r|^{-a_{k}}
\label{eq36}
\eeqa 
where, $\alpha, \gamma^\prime, a_{k}= 1.8, 1.2, 0.63$ respectively are the critical exponents.
\section{Results and Discussion}
Now we discuss the effects of CEP  on the EoS and subsequently on the $\Snn$.
We have evaluated the $\Snn$ by including the effects of CEP
through (i) the  EoS and (ii) the critical behaviour of various 
transport coefficients as discussed above. 
As mentioned in section~\ref{QCDEoS}  we assume that the CEP 
is located at $(\mu,T)=(367,154)$ MeV in the QCD phase diagram. 
This means that the entropy density (first order derivative of free energy) 
changes continuously for $\mu<367$ MeV and $T>154$ MeV reflecting the 
change over to crossover from first order transition (characterized by a discontinuity 
in entropy density) for $\mu>367$ MeV and $T<154$ MeV~\cite{hasan}.

\begin{figure}[h]
\centering
\includegraphics[width=8.5cm]{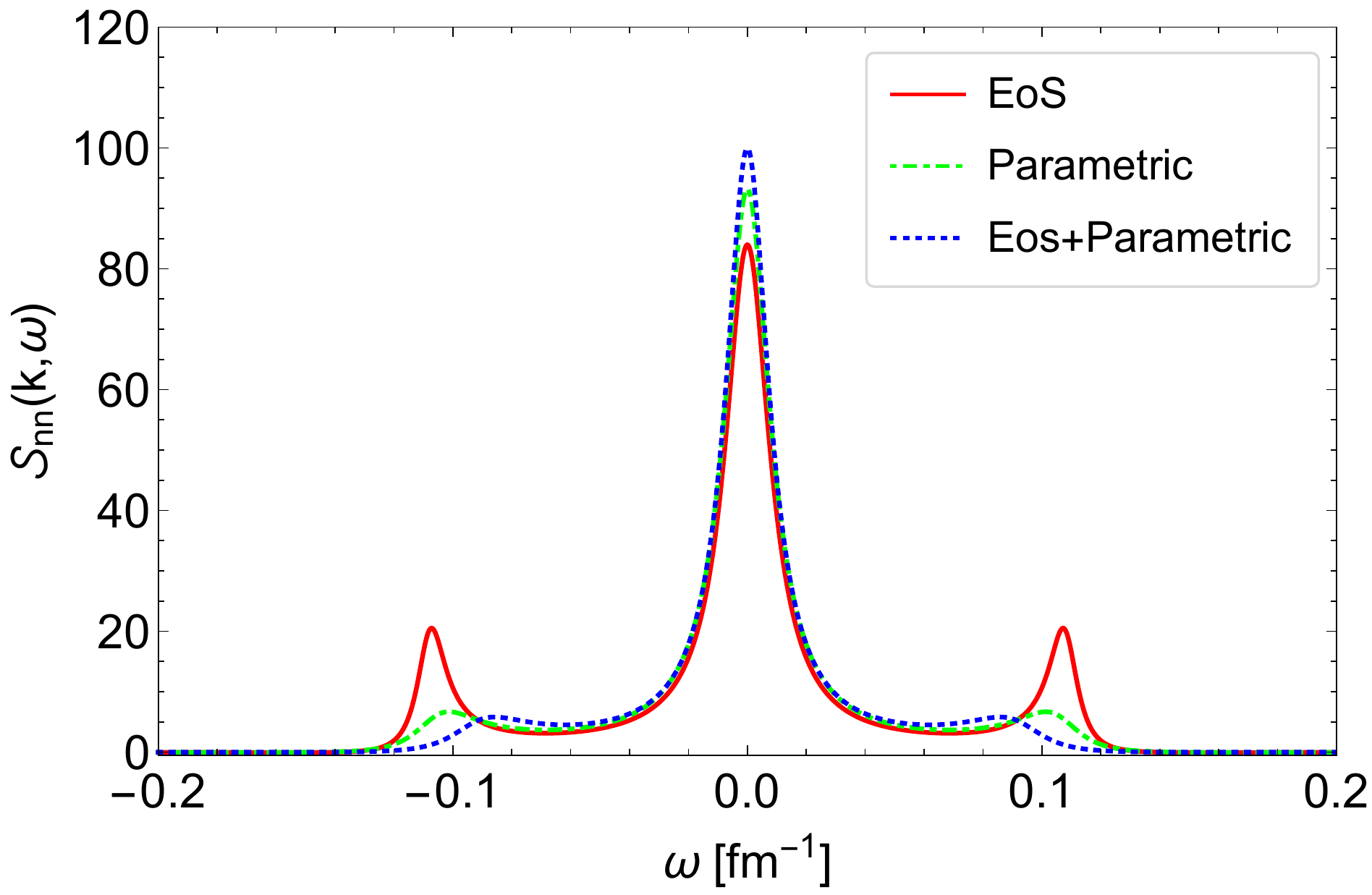}	
\caption{Variation of $\Snn$ with $\omega$ for point ($r=0.2$) and $k=0.1$ fm$^{-1}$. 
Red line shows the effects of EoS with $\eta/s= \zeta/s =\kappa/s=1/4\pi$.  
The green line is obtained for parametric forms of  thermodynamic quantities. And blue line is  drawn with both effects.
}
\label{fig1}
\end{figure}
\begin{figure}[h]
\centering
\includegraphics[width=8.5cm]{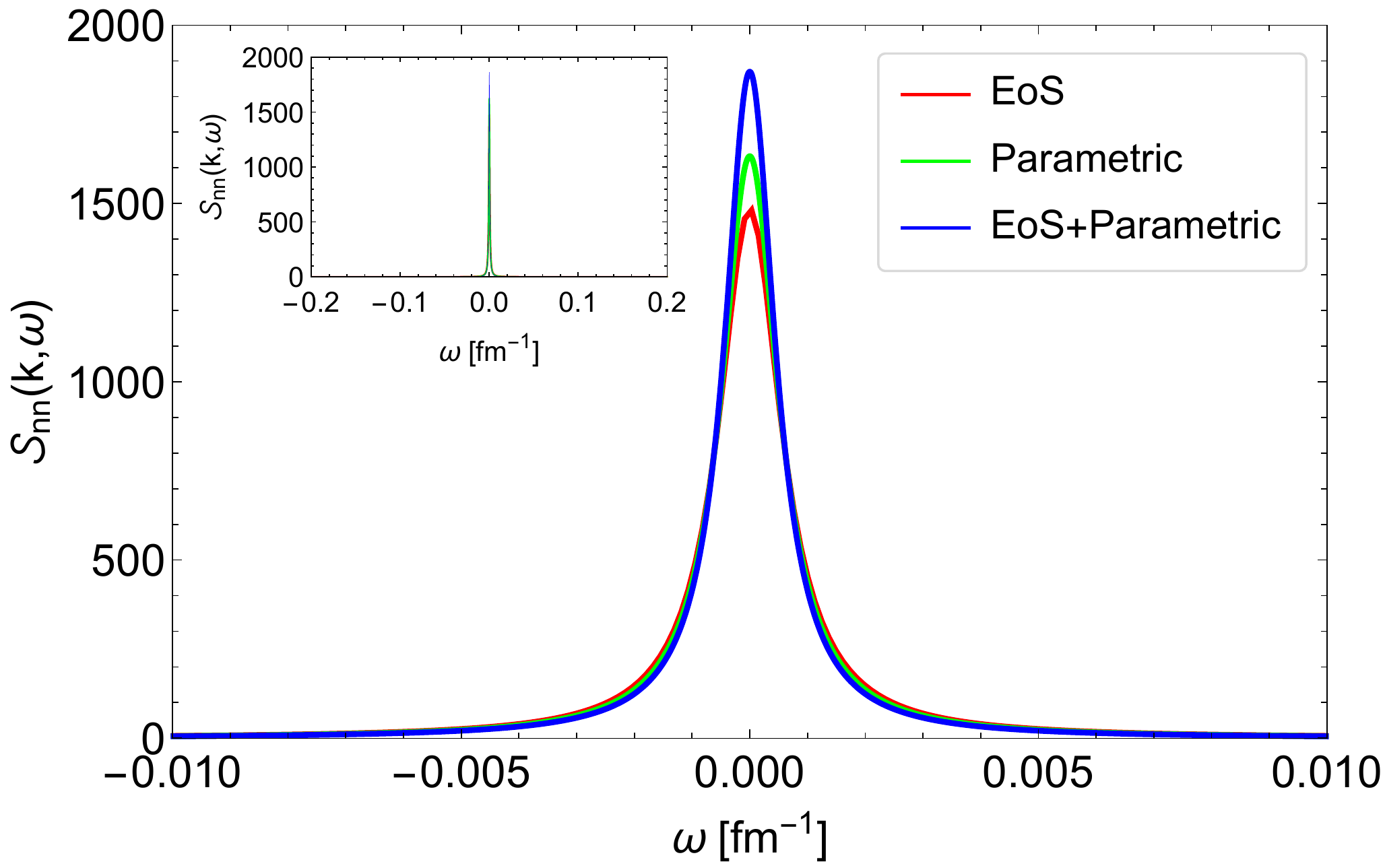}	
\caption{$\Snn$ near the critical point for ($r=0.01$) and $k=0.1$ fm$^{-1}$. 
The red curve represents the effects of EoS and  $\eta/s= \zeta/s =\kappa/s=1/4\pi$.
The green line is  obtained
by using the scaling hypothesis of thermodynamic variables near CEP and both effects are shown in blue curve. Inset plot is for the broader range in $\omega$ ($-0.2\le\omega\le 0.2$).}
\label{fig2}
\end{figure}
Fig.~\ref{fig1} displays the variation of $\Snn$ with $\omega$  when the
system is away from CEP represented by $r=0.2$. Results with the effects of 
(i), (ii) as well as with the combined effects of (i)+(ii) are shown.  
To investigate the effects of EoS we keep  
the transport coefficients finite with values,  $\eta/s, \zeta/s, \kappa/s=1/4\pi$. 
The relevant coupling  and relaxation coefficients are estimated from Eq.\ref{eq14}. 
We observe that for (i) there are three distinct peaks (red line).
The central one is large and called the  R-peak originates from the entropy fluctuation
at constant pressure {\it i.e.} due to thermal
fluctuation. The symmetric doublet about the R-peak are the B-peaks. The B-peaks
originates from pressure fluctuation at constant entropy which is connected to sound wave. 
We find that the B-peaks are weaker than the R-peaks.
(ii) We exclude the effects of EoS  to investigate the effects of transport coefficients only
near the  CEP via their critical exponents  
in evaluating $\Snn$.
The dotted green curve in Fig.~\ref{fig1} depicts the $\Snn$ obtained by
using the scaling laws i.e when the parametric forms of the  transport coefficients and 
response functions mentioned in Eq.\ref{eq36} are used. 
The B-peaks in scenario (ii) is sub-dominant, almost like  a broad shoulder.  The combined
effects is depicted by the blue dotted line in Fig~\ref{fig1}. It is observed 
that the combined effects make the B-peaks sub-dominant. 

Closer to the CEP, however, the scenario changes drastically. The 
$\Snn$ near the critical point ($r=0.01$) shows
only the R-peak with height increased by more than an order of magnitude and
the B-peaks vanish due to absorption of sound waves near CEP (Fig.\ref{fig2}). 
The vanishing of B-peaks can be understood from the fact that in leading order 
the Brillouin frequencies, $\pm\omega_B\sim c_s k$ where $c_s$ is the velocity of sound. 
The effects of EoS is seen to reduce the height of the R-peak (red line). 
\begin{figure}[h]
\centering
\includegraphics[width=8.5cm]{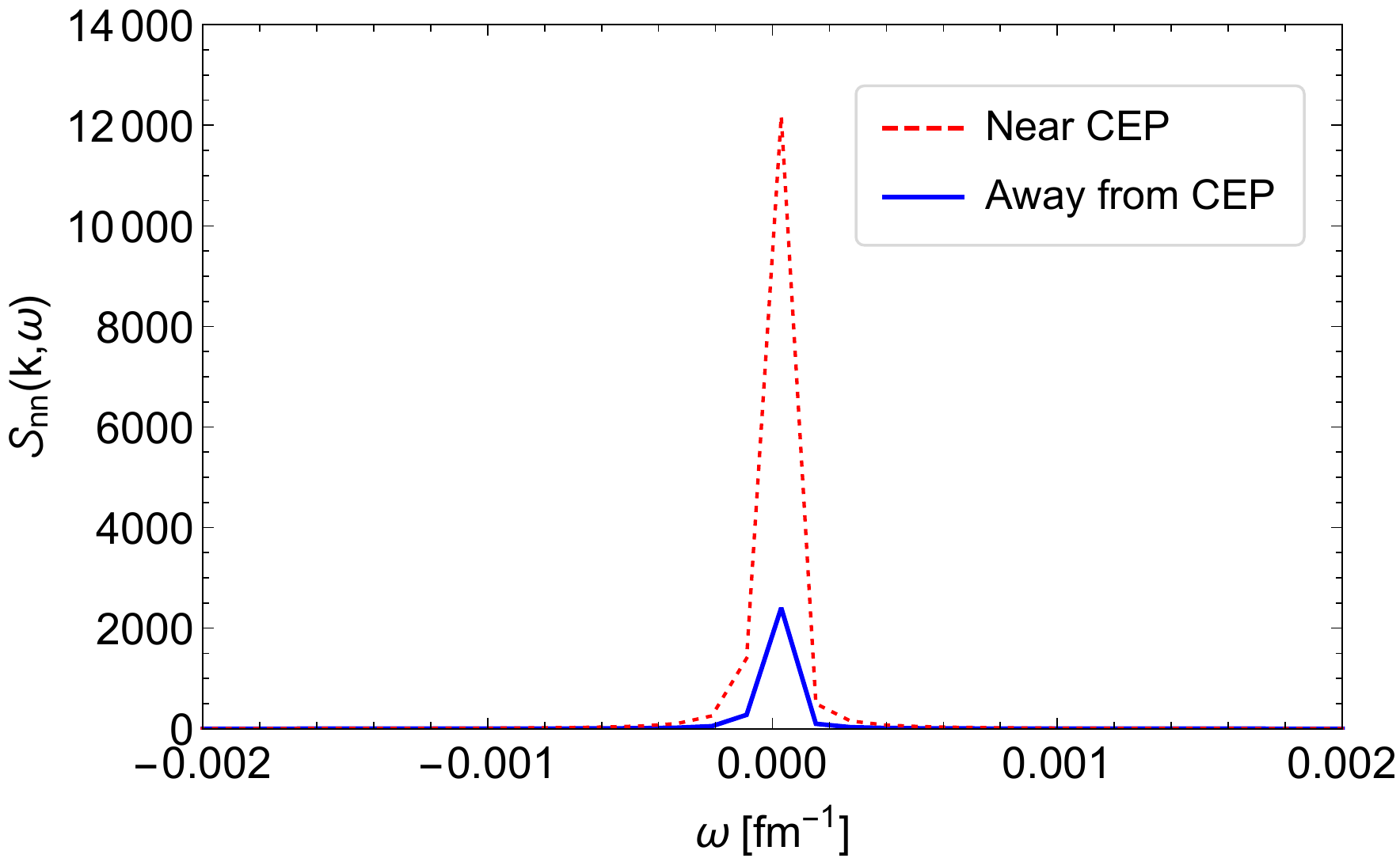}	
\caption{Variation of $\Snn$ with $\omega$ for $r=0.01$ (red-line)
and 0.2 (blue-line) corresponding to the situation
of the system near and away from  the CEP
respectively. The value of $k$ is $0.02$ fm$^{-1}$ and $\eta/s= \zeta/s =\kappa/s=1/4\pi$.}
\label{fig3}
\end{figure}

The $\Snn$ as a function of $\omega$ for smaller $k$ ($k=0.02$ fm$^{-1}$) is
plotted in Fig.\ref{fig3} when only the effects of EoS is considered. 
The red (blue) line corresponds to results close to the CEP with $r=0.01$  
(away from CEP with $r=0.2$, B-peaks are not visible because of the scale chosen 
along $\omega$ axis). 
We observe that at smaller $k$ the value of R-peak gets larger as well as sharper.  
A comparison with results shown in Fig.\ref{fig1} reveals that near CEP, 
the  B-peaks vanish and the R-peak gets sharper and larger. 
\begin{figure}[h]
\centering
\includegraphics[width=8.5cm]{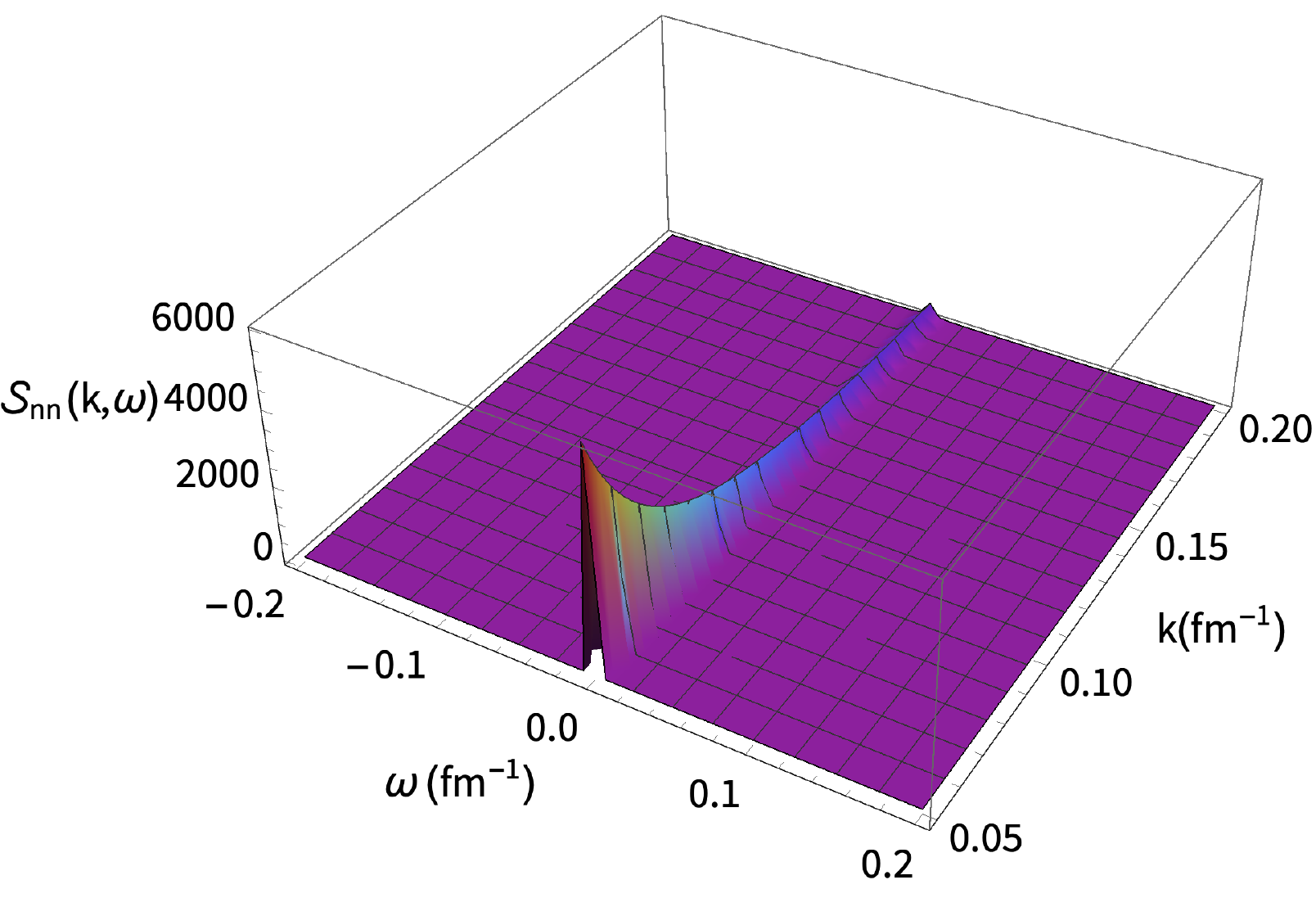}	
\caption{Variation of $\Snn$ with $\omega$ and $k$ near the CEP ($r=0.01$)
for  $\eta/s= \zeta/s =\kappa/s=1/4\pi$}
\label{fig4}
\end{figure}
\begin{figure}[h]
\centering
\includegraphics[width=8.5cm]{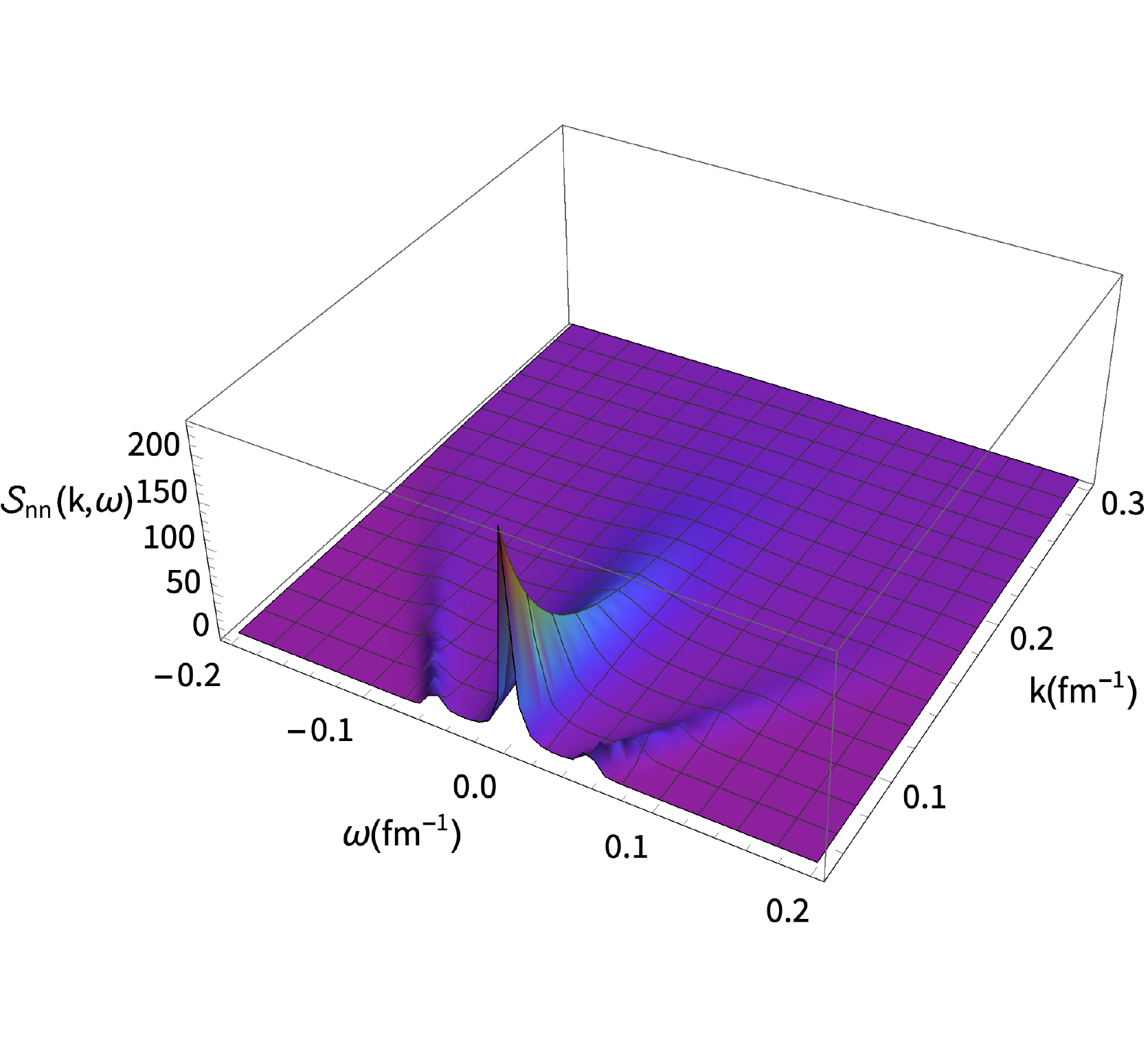}	
\caption{Same as Fig.~\ref{fig4} for $r=0.2$ {\it i.e.} the system is
away from CEP.}
\label{fig5}
\end{figure}
\begin{figure}[h]
\centering
\includegraphics[width=8.5cm]{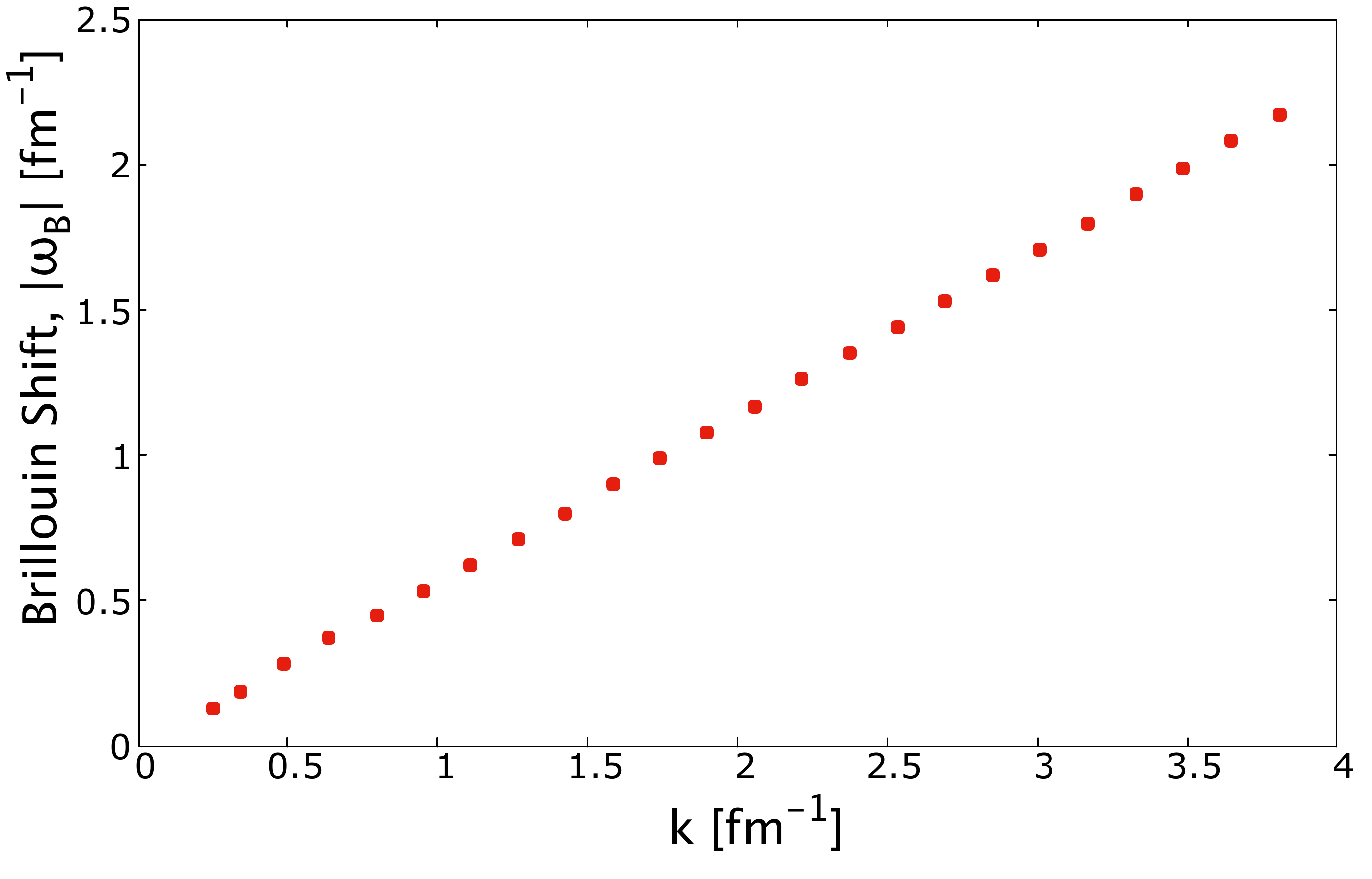}	
\caption{Variation of the Brillouin frequency, $\omega_B$ with $k$ 
when the system is away from CEP ($r=0.2$) with $\eta/s= \zeta/s =\kappa/s=1/4\pi$.}
\label{fig6}
\end{figure}
Motivated by the above results shown in Figs.~\ref{fig2} and \ref{fig3} {\it i.e.} by observing the sensitivity of
the results on the $k$ values we investigate the behaviour of spectral function for different 
hydrodynamic modes ($k-$modes). We consider the variation of $\Snn$ with  $k$ and $\omega$ 
near the CEP in Fig.\ref{fig4}. We observe that for all values of $k$, the B-peaks merge 
with the R-peak. 
The height of the peak is maximum in the neighbourhood of $k\rightarrow 0$. 
In Fig.\ref{fig5} the $\Snn$ is plotted against $\omega$ and $k$ when the system is away from CEP. 
We observe that the B-peaks shift away from R-peak with increase in $k$. 
This is better reflected  in Fig.\ref{fig6}, where the Brillouin frequency is plotted
against $k$. A linear variation of $\omega$ with $k$ is obtained for 
given values of $T$ and $\mu$.  The slope of the line is found to be 0.28 close to the velocity
of sound $c_s=0.25$  obtained from Eq.~\ref{cs2} at the same value of temperature and chemical potential. 
Therefore, we find that  the B-peaks move toward the R-peak 
as the system approach the CEP and ultimately merge with R-peak at
the CEP.  This  suggests that the speed of propagation of all the hydrodynamic 
modes vanish at the CEP.  This is consistent with the finding that the speed of 
sound reaches the minimum at QCD the critical point. 
The positions of the B-peak  depends on $k$ when the system is away from CEP which
indicates that the different k-modes travel with different speeds in the medium.  

This nature of mode dependent speed of sound also signifies that the attenuation of the Mach-cone can not be 
interpreted as the  unique effect of critical point as suggested in Ref.~\cite{kunihiro}. Rather mode dependent 
speed of propagation at points away from the CEP and vanishing speed of sound at CEP should 
be taken into account to properly interpret  the attenuation of Mach-cone structure in the particle correlations.


\section{summary and discussions}

The correlation of density fluctuation near QCD critical point has been studied by using linearized perturbative
equations obtained from Israel-Stewart hydrodynamics. The spectral structure, $\Snn$ of the density fluctuation  
has been derived rigorously by keeping all the relevant transport coefficients and response functions non-zero.  
  The effects of EoS and the critical behaviour of various transport coefficients on  $\Snn$ have been investigated.

 
The mode dependent propagation speed away from CEP suggests that careful  look is needed for interpreting the 
suppression of Mach cone structure as a  signature of CEP. These findings suggest that beam energy dependence of  
the correlation of number density fluctuation of produced particles in RHIC-E has the potential 
to carry the signature of critical point.  

In condensed matter physics the  effects of critical point
have been investigated by measuring the intensity of the
light scattered from the system.  In contrast to this 
no such external probe is available for detecting CEP in QCD system. 
However, several possibilities for the detection of CEP in QCD have been
discussed  in the literature. 
In condensed matter system the phenomenon of critical opalescence
at the CEP is considered as  signal of large density 
fluctuations~\cite{Stanley}.
The possibility of detecting the phenomenon of QCD opalescence by measuring  the
suppression of hadronic spectra in RHIC-E ($R_{AA}$) has been indicated in  
~\cite{csorgo}.
The $R_{AA}$ can be used to estimate the opacity factor, $\K$ as: 
\begin{equation}
\K=-\frac{ln(R_{AA})}{R_{HBT}}
\end{equation}
where $R_{HBT}$ is the Hanbury-Brown Twiss radius of the system. 

The Fourier coefficients of the azimuthal distributions of 
particles can be used to understand various
properties of the matter produced in RHIC-E.  
The coefficient of $cos3\phi$ (triangular flow)
sheds light on the the initial fluctuations, similarly,
the coefficients of $cos2\phi$ (elliptic flow)
can be used to discern the EoS of the system.
Near the critical point the order of
the harmonics vary as $\sim 1/\lambda_c$ where $\lambda_c$ is the
wavelength of the perturbation in pressure (sound) which diverges 
at the CEP and hence all the harmonics will vanish~\cite{hasan}.
However, the experimentally measured  spectra contains
contribution from all the space-time points 
{\it i.e.} from all possible values of $T$ and $\mu$
not only from CEP.  Therefore, even if the system 
passes through the CEP the Fourier coefficients may not vanish, 
but the CEP may depress them.

In Ref.~\cite{Kapusta} the mode-mode coupling theory has been 
used to study the existence and detection of CEP.
It is comprehensively shown in ~\cite{Kapusta} that the thermal conductivity
at CEP diverges which induces a sharp change in the
two particle correlation of fluctuations in rapidity space.
Detection of such modifications in the correlation
function may confirm the existence of CEP.
The suppression of fluctuations in temperature ($\Delta T$) and baryonic chemical potential ($\Delta\mu$)
due to the divergence of thermodynamic 
response functions at CEP
can signal the presence of CEP  
\cite{Stephanov:1998dy,Stephanov:1999zu}. The  suppression in $\Delta T$ and $\Delta\mu$ will be reflected through the 
transverse momentum spectra of hadrons and proton to pion ratio respectively.

In reality, the possibility of the trajectories passing through 
the critical point is remote. This limits 
the fluctuations near CEP. These fluctuations will remain  
out of equilibrium due to critical slowing down~\cite{stephanov3}.
The  creation of defects  due to CEP in QCD
like cosmology ~\cite{kibble} and condensed matter~\cite{zurek} systems
and their detection will be extremely exciting.
The appearance of Kibble-Zurek length scale and its connection with
spatial correlations have been studied~\cite{yakamatsu}.
The measurement of enhancement of non-flow correlations in presence of CEP
as function of $n/s$ can be used to detect
the CEP~\cite{yakamatsu}.

Rigorously speaking fluid dynamics works in the region where
the condition, $k<<q$ is satisfied where  $q$ is the 
inverse of correlation length which can be expresses as $q=q_o\,r^{\nu}$
$q_o$ is a constant and $\nu$ is the critical index with a numerical value 
$\nu=0.73\pm 0.02$~\cite{krg}.
This fundamental assumption becomes inoperative 
since $\xi$ diverges ($q\rightarrow 0$) at the CEP. 
However, there will be a region in the neighbourhood of CEP  
where the predictions of fluid dynamics may be useful. 
Following a procedure similar to condensed matter system~\cite{Stanley} we can
write, $k<<q_o r^{\nu}$ for the validity of the hydrodynamics, which implies,
\begin{equation}
T>T_c\left[1+\left(\frac{k}{q_o}\right)^{1/\nu}\right]
\end{equation}
Since hydrodynamics is an effective theory for soft physics (large wavelength  or small wave vector, $k$),
fluid dynamics can be applied in the neighbourhood of CEP but becomes invalid at CEP because
of the divergence of the correlation length. 
The response of the trajectories in the neighbourhood of CEP in the $\mu-T$ plane   
to the initial conditions away from equilibrium 
has been investigated in Ref.~\cite{tdore}. 

In a realistic scenario the matter formed in RHIC-E 
expands hydrodynamically which is not taken into consideration
in the present work. The effects of the CEP in  
(3+1) dimensionally expanding system within the 
purview of second order viscous hydrodynamics is currently under
investigation~\cite{sksingh}.

The physics of hadronic matter under extreme conditions of temperatures or densities
and the QCD phase transition has been considered as the condensed matter
physics of elementary particles~\cite{cmp} where the relevant microscopic interaction is controlled
by non-abelian gauge theory in contrast to the condensed matter physics governed 
by abelian gauge theory. In condensed matter physics the theoretical results on dynamical
spectral structure can be compared with the experimentally measured intensity of 
scattered of light. It has been shown that experimentally 
measured widths of R and B-peaks 
and one of the transport coefficients  (among $\eta$, $\zeta$ and $\kappa$)   
as inputs will lead to the determination of  the other two transport coefficients. 
Moreover, the integrated
intensities of R and B-peaks has been connected to the 
Landau-Placzek ratio,  $C_p/C_v-1=\gamma-1$. 
Identification of appropriate probes analogous to light in abelian system will go a long 
way to reveal the physics QCD matter near CEP. 

\section{Appendix A}
In this appendix the expression for the dynamical spectral structure, $\Snn$  derived by considering contributions 
up to second order in transport coefficients (i.e $\eta^{2},\zeta^{2},\kappa^{2},\eta \zeta, \eta \kappa,\zeta\kappa$)
has been provided.  The coupling and relaxation coefficients ($\alpha_{0},\alpha_{1},\beta_{0},\beta_{1},\beta_{2}$) 
have been taken non-zero in obtaining  the results displayed in the text, but have been taken 
as zero in the following to avoid  a more lengthy and complex expressions.
\beqa
\mathcal{S^\prime}_{nn}(\vec{k},\omega)&=&k^2 n_0 \Bigg[ \omega ^2 n_0 T^{2}\big(\frac{\text{$\pd $s}}
{\text{$\pd $T}}\big)_n \Big\{h_0 \kappa  \omega+
\big(\frac{\text{$\pd $p}}{\text{$\pd $T}}\big)_n  k^2 \big( 
\zeta+\frac{4}{3} \eta  \big)-k^{2}T \kappa  \omega ^2+h_{0}k \kappa
\big(\frac{\text{$\pd $s}}{\text{$\pd $n}}\big)_T  \nn\\
&&+k^2 \kappa  ( \zeta +\frac{4}{3}\eta )- 
T^{2} \big(\frac{\text{$\pd $p}}{\text{$\pd $T}}\big)_n^2\Big\}+k^2 \kappa h_{0}  \Big\{- h_0T \big(\frac{\text{$\pd
   $p}}{\text{$\pd $T}}\big)_n+k^2 \kappa  (\zeta+\frac{4}{3} \eta  )+T^{2} \big(\frac{\text{$\pd $p}}{\text{$\pd
   $T}}\big)_n^2\Big\}\nn\\
   &&+n_0^{2}T^{2} \big(\frac{\text{$\pd $s}}{\text{$\pd $T}}\big)_n \big(\frac{\text{$\pd $s}}{\text{$\pd $n}}\big)_T
   \Big\{\big(\frac{\text{$\pd $p}}{\text{$\pd $T}}\big)_n k ( \zeta +\frac{4}{3} \eta )+T^{2} \kappa  \omega ^2- h_0 \kappa \Big\}+k^4 n_{0} h_{0} (\zeta +\frac{4}{3} \eta )^{2}\nn\\
   &&+h_{0}\big(\frac{\text{$\pd $p}}{\text{$\pd $n}}\big)_T \Big\{n_0 T^2 \omega^{2}
    \Big(k^2 (\zeta +\frac{4}{3} \eta )+ \kappa  \omega ^2- k^2 \kappa
  \Big) \Big\}+ h_{0}^{2} \kappa T k^{2}  \big(\frac{\text{$\pd $p}}{\text{$\pd $T}}\big)_n\nn\\
  &&- n_0h_{0} \kappa T^{2} k^{2}\omega ^2 \big(\frac{\text{$\pd
   $s}}{\text{$\pd $n}}\big)_T\Bigg] \Bigg / \Bigg[ h_0^2 \Big\{ k^4 \kappa^2 \omega ^4+n_0^2 T^2 \omega ^4\big(\frac{\text{$\pd $s}}{\text{$\pd $T}}\big)_n\Big\}+h_0 k^2
   \omega ^2  \kappa^2T^{2}\big(\frac{\text{$\pd $s}}{\text{$\pd $T}}\big)_n  \nn\\
   &&\Big\{  \omega ^4 -2n_{0} \big(\frac{\text{$\pd
   $p}}{\text{$\pd $n}}\big)_T\Big\}-2 k^2 n_0^2 T^{2} \kappa^2 \omega ^2\big( \frac{\text{$\pd $s}}{\text{$\pd $T}}\big)_n^{2}-2n_{0}h_{0}k^2 T^{2} \kappa^2
   \omega ^2 \big(\frac{\text{$\pd $p}}{\text{$\pd $T}}\big)_n\nn\\
   &&-2n_0^3 T^2k^{4} \omega ^2 \big(\frac{\text{$\pd $s}}{\text{$\pd $T}}\big)_n \Big\{
   \big(\frac{\text{$\pd $p}}{\text{$\pd $T}}\big)_n \big(\frac{\text{$\pd $s}}{\text{$\pd $n}}\big)_T+ \big(\frac{\text{$\pd $s}}{\text{$\pd
   $T}}\big)_n \big(\frac{\text{$\pd $p}}{\text{$\pd $n}}\big)_T\Big\}+k^4 n_0^4 T^2 \omega ^2 \big(\frac{\text{$\pd
   $s}}{\text{$\pd $n}}\big)_T^2  \big(\frac{\text{$\pd $p}}{\text{$\pd $n}}\big)_T^2\nn\\
   &&+2k^2
   n_0 T \kappa  \omega ^2 \big(\frac{\text{$\pd $p}}{\text{$\pd $T}}\big)_n \Big\{ k^4 \kappa  \big(\frac{\text{$\pd $p}}{\text{$\pd $n}}\big)_T+T \omega ^2 \big(\frac{\text{$\pd $s}}{\text{$\pd $n}}\big)_T k^2(  \zeta+\frac{4}{3} \eta  )-T^{3} \kappa  \omega
   ^2\Big\}\nn\\
   &&+n_0^2 T^{2}\Big\{ h_{0}k^{2} \kappa^2 \big(\frac{\text{$\pd $p}}{\text{$\pd $n}}\big)_T^2+k^{2}T^2 \kappa^2 \omega ^4-2 k^4 T \kappa \omega ^4
    \big(\frac{\text{$\pd $p}}{\text{$\pd $n}}\big)_T^{2}+2k^2 h_{0}
   \kappa  \omega ^2 ( \zeta +\frac{4}{3} \eta )\nn\\
    &&+\big(\frac{\text{$\pd $s}}{\text{$\pd $T}}\big)_n^2 k^4 ( \zeta+\frac{4}{3}\eta )^{2}+2k^4T \kappa  \big(\frac{\text{$\pd $p}}{\text{$\pd $T}}\big)_n
   \big(\frac{\text{$\pd $s}}{\text{$\pd $n}}\big)_T  (\zeta +\frac{4}{3} \eta )\Big\}\Bigg]
   \big< \delta n(\vec{k},0)\delta n(\vec{k},0)\big>
\eeqa
where
\begin{equation}
\mathcal{S}_{nn}(\vec{k},\omega)=\frac{\mathcal{S^\prime}_{nn}(\vec{k},\omega)}
   {\big< \delta n(\vec{k},0)\delta n(\vec{k},0)\big>}
\end{equation}
The expression for $\mathcal{S}_{nn}(\vec{k},\omega)$ contain derivatives of several thermodynamics quantities. In this appendix
we recast these derivatives
in terms of response functions like: isothermal and adiabatic compressibilities  ($\kappa_T$ and $\kappa_s$), 
specific heats ($C_p$ and $C_v$), baryon number susceptibility ($\chi_B$) and velocity of sound ($c_s$), etc. The 
baryon number density ($n$) and the entropy density ($s$) can be written as: 
\beqa
n=\Big(\frac{\partial p}{\partial \mu}\Big)_{T};\,\,\,\,\, s=\Big(\frac{\partial p}{\partial T}\Big)_{\mu}
\eeqa
Baryon number susceptibility, isothermal compressibility and adiabatic compressibility are given by,
\beqa
\chi_{B}=\Big(\frac{\partial n}{\partial \mu}\Big)_{T}; \kappa_{T}=\frac{1}{n}\Big(\frac{\partial n}{\partial p}\Big)_{T}; \kappa_{s}=\frac{1}{n}\Big(\frac{\partial n}{\partial p}\Big)_{s}
\eeqa
Specific heats  can be expressed as:
\beqa
C_{p}=T\Big(\frac{\partial s}{\partial T}\Big)_{p}; C_{v}&=&T\Big(\frac{\partial s}{\partial T}\Big)_{V}=T\Big(\frac{\partial s}{\partial T}\Big)_{n}\nn\\
&=&\Big(\frac{\partial \epsilon}{\partial T}\Big)_{V}=\Big(\frac{\partial \epsilon}{\partial T}\Big)_{n}
\eeqa
Now we write down the expression for partial derivatives, $(\frac{\partial p}{\partial T})_{n}, 
(\frac{\partial p}{\partial n})_{T}, (\frac{\partial \epsilon}{\partial T})_{n}$ and
$\Big(\frac{\partial \epsilon}{\partial n}\Big)_{T}$ below. 
$(\frac{\partial p}{\partial T})_{n}$ can be evaluated as: 
\beqa
\Big(\frac{\partial p}{\partial T}\Big)_{n}&=&\frac{\pd (p, n)}{\pd (T,n)}\nn\\
&=&\frac{\pd (p, n)}{\pd (T, p)}\frac{\pd (T, p)}{\pd (s, p)}\frac{\pd (s, p)}{\pd (s, \epsilon)}\frac{\pd (s, \epsilon)}{\pd (s, n)}\frac{\pd (s, n)}{\pd (T, n)} \nn\\
&=& \Big[-\Big(\frac{\partial n}{\partial T}\Big)_{p}\Big]\Big(\frac{\partial T}{\partial s}\Big)_{p}\Big(\frac{\partial p}{\partial \epsilon}\Big)_{s}\Big(\frac{\partial \epsilon}{\partial n}\Big)_{s}\Big(\frac{\partial s}{\partial T}\Big)_{n} \nn\\
&=& n\alpha_{p}\frac{T}{C_{p}}c^{2}_{s}\Big(\frac{\partial \epsilon}{\partial n}\Big)_{s}\frac{C_{v}}{T}\nonumber\\
&=&nc^{2}_{s}\alpha_{p} \frac{C_{v}}{C_{p}}\Big(\frac{\partial \epsilon}{\partial n}\Big)_{s}
\eeqa
Using the relation,
\beqa
d\epsilon=Tds+\mu dn \,\,\,\,\, \text{and } \mu=\Big(\frac{\partial \epsilon}{\partial n}\Big)_{s}
\eeqa
we write: 
\beqa
\Big(\frac{\partial p}{\partial T}\Big)_{n}=\mu nc^{2}_{s}\alpha_{p} \frac{C_{v}}{C_{p}}
\eeqa 

Next we consider $\Big(\frac{\partial p}{\partial n}\Big)_{T}$:
\beqa
{\Big(\frac{\partial p}{\partial n}\Big)_{T}= \frac{1}{n\kappa_{T}}}
\eeqa
\beqa
\Big(\frac{\partial p}{\partial n}\Big)_{T}&=&\frac{\pd (p, T)}{\pd (n, T)}
=\frac{\pd (p, T)}{\pd (p, s)}\frac{\pd (p, s)}{\pd (\epsilon, s)}\frac{\pd (\epsilon, s)}{\pd (n,s)}\frac{\pd (n, s)}{\pd (n, T)}\nn\\
&=&\Big(\frac{\partial T}{\partial s}\Big)_{p}\Big(\frac{\partial p}{\partial \epsilon}\Big)_{s}\Big(\frac{\partial \epsilon}{\partial n}\Big)_{s}\Big(\frac{\partial s}{\partial T}\Big)_{n}\nn\\
&=& \frac{T}{C_{p}}c^{2}_{s}\Big(\frac{\partial \epsilon}{\partial n}\Big)_{s}\frac{C_{v}}{T}
\\&=&\mu c^{2}_{s} \frac{C_{v}}{C_{p}}
\eeqa
 
The factor, $\Big(\frac{\partial s}{\partial T}\Big)_{n}$ can be written as:
\beqa
\Big(\frac{\partial s}{\partial T}\Big)_{n}=\frac{1}{T_{0}}\Big(\frac{T_{0}\partial s}{\partial T}\Big)_{n}=C_{v}
\eeqa
For fixed net baryon number,  $c_n$ can be written as $c_{n}=C_{v}$.
Therefore, 
\beqa
{\Big(\frac{\partial \epsilon}{\partial T}\Big)_{n}=C_{v}}
\eeqa
We  evaluate the derivative   $\Big(\frac{\partial \epsilon}{\partial n}\Big)_{T}$  as
\beqa
\Big(\frac{\partial s}{\partial n}\Big)_{T}&=&-\Big(\frac{\partial s}{\partial T}\Big)_{n}\Big(\frac{\partial T}{\partial n}\Big)_{s}=-\frac{1}{T_{0}}\Big(\frac{T\partial s}{\partial T}\Big)_{n}\Big(\frac{\partial T}{\partial n}\Big)_{s}\nn\\
&=& -\frac{C_{v}}{T_{0}}\frac{1}{n_{0}}\Big[n_{0}\Big(\frac{\partial T}{\partial n}\Big)_{s}\Big]= \frac{C_{v}}{n_{0}T_{0}\alpha_{s}}
\eeqa
The velocity of sound is given by:
\beqa
c_s^2=\Big(\frac{\pd p}{\pd \epsilon}\Big)_{s/n}&=& \frac{nd\mu+sdT}{\mu dn+Tds}\nn\\
&=& \frac{nF dT+sdT}{\mu (\frac{\partial n}{\partial T})_\mu  dT+ \mu(\frac{\partial n}{\partial \mu})_T d\mu+T(\frac{\partial s}{\partial T})_\mu  dT+ T(\frac{\partial s}{\partial \mu})_T d\mu}\nn\\
&=&  \frac{nF dT+sdT}{\mu (\frac{\partial n}{\partial T})_\mu  dT+ \mu F(\frac{\partial n}{\partial \mu})_T dT+T(\frac{\partial s}{\partial T})_\mu  dT+ TF(\frac{\partial s}{\partial \mu})_T dT} \nn\\
&=&  \frac{nF+s}{\mu (\frac{\partial n}{\partial T})_\mu + \mu F(\frac{\partial n}{\partial \mu})_T+T(\frac{\partial s}{\partial T})_\mu+ TF(\frac{\partial s}{\partial \mu})_T}
\label{cs2}
\eeqa
where,
\beqa
F&=& \frac{(\frac{\partial s}{\partial T})_\mu- 
\frac{s}{n}(\frac{\pd n}{\pd T})_{\mu}}{\frac{s}{n}(\frac{\pd n}{\pd \mu})_{T}
-(\frac{\partial s}{\partial \mu})_T}\nn\\
\eeqa
\section{Acknowledgement}
M.R. is supported by Department of Atomic Energy (DAE), Govt. of India.
The work of AB is  supported by Alexander von Humboldt (AvH) foundation and Federal Ministry
of Education and Research (Germany)  through Research Group Linkage programme.  AB also thanks 
Purnendu Chakraborty, Sourin Mukhopadhyay and Soumen Datta for fruitful discussions.

\end{document}